# EXPERIMENTAL CHARACTERIZATION OF THE STATIC BEHAVIOUR OF MICROCANTILEVERS ELECTROSTATICALLY ACTUATED

*Alberto Ballestra[1], Eugenio Brusa[2], Mircea Georghe Munteanu[2], Aurelio Somà[1]*

[1] Laboratory of Microsystems, Department of Mechanics, Politecnico di Torino,
C.so Duca degli Abruzzi, 24 – 10129 Torino, Italy; alberto.ballestra@polito.it, aurelio.soma@polito.it

[2] Department of Electrical, Management and Mechanical Engineering, Università di Udine,
via delle Scienze, 208 – 33100 Udine, Italy; eugenio.brusa@uniud.it, munteanu@uniud.it

**ABSTRACT**

This paper concerns the experimental validation of some mathematical models previously developed by the authors, to predict the static behaviour of microelectrostatic actuators, basically free-clamped microbeams. This layout is currently used in RF-MEMS design operation or even in material testing at microscale. The analysis investigates preliminarily the static behaviour of a set of microcantilevers bending in-plane. This investigation is aimed to distinguish the geometrical linear behaviour, exhibited under small displacement assumption, from the geometrical nonlinearity, caused by large deflection. The applied electromechanical force, which nonlinearly depends on displacement, charge and voltage, is predicted by a coupled-field approach, based on numerical methods and herewith experimentally validated, by means of a Fogale Zoomsurf 3D. Model performance is evaluated on pull-in prediction and on the curve displacement vs. voltage. In fact, FEM nonlinear solution performed by a coupled-field approach, available on commercial codes, and by a FEM non-incremental approach are compared with linear solution, for different values of the design parameters.

## 1. INTRODUCTION

In microsystem mechanical design cantilever beams are currently widely used, as basic components in microsensors, microswitches and RF-MEMS as well as in experimental micromechanics, whose goal is characterizing the materials mechanical properties and strength, at microscale [1,2,3,4]. The latter aspects motivate the implementation of efficient numerical models to predict the electromechanical behaviours of such microdevices, under the actuation of the electric field, as stand-alone systems or better as structural components of assembled parts, as recent DTIP Conferences showed during the last years, like in [5-8]. Model validation is currently performed not only to verify the effectiveness of proposed analytical, numerical and even compact approaches [9, 10], but to define the model sensitivity on the uncertainties about the actual values of the design parameters and of materials properties, whose measurement is often fairly difficult. A couple of targets appear currently challenging for structural micromechatronics. An assessment of accurate coupled-field models and numerical solutions shall allow a coherent interpretation of the specimen response in all the experimental procedures, currently performed and aimed to characterize both the materials and the MEMS layouts [3, 4, 11, 12, 13]. Moreover, to build effective numerical simulators, able to predict the coupled behaviour of MEMS within the whole electronic circuit, only a validation of each single model included in a hierarchical approach will allow satisfying the requirement [9]. This paper contributes to the above mentioned tasks, by investigating the effectiveness and the computational performance of the numerical models proposed in [1, 14-18], dealing with the static behaviour of microcantilevers. Moreover, the above mentioned models have to be even used in dynamic analysis algorithms, when geometrical nonlinearity has to be added to the effects of nonlinear electromechanical [16-18].

## 2. THE EXPERIMENTAL SET-UP

### 2.1. Microcantilevers with in-plane bending

A first group of specimens including free-clamped microcantilevers was designed and built, according to the design rules and the process constraints imposed by microfabrication, followed by STMicroelectronics (Cornaredo, Italy). Process "Thelma" allows a gradual growth of thick polysilicon layers, being suitable to fabricate cantilever beams, for which the bending deflection occurs in-plane, with respect to the reference plane of the wafer (Fig.1). This approach was followed to validate the developed models, by means of the experimental measures performed by Fogale Zoomsurf 3D [19]. All microspecimens consist of a massive electrode where is clamped a thin microbeam, bending across the gap towards a massive counter-electrode.

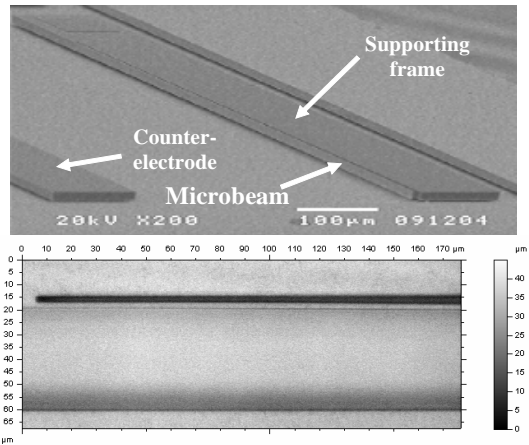

*Figure 1: Microcantilever specimen in-plane bending*

Microcantilever is a part of a wider structure, equipped by a connection pad. Thickness and length are measured in the plane of the wafer, while the width is measured along the direction orthogonal to the wafer plane. The electric potential is imposed on the electrode, where microbeam is clamped, through the pad and to the connection pad of the counter-electrode, thus applying the voltage through the gap. To perform a parametric investigation several lengths were foreseen, as well as different values of gaps were obtained. Dimensions are listed in Table 1. The material of specimens is $PoCl_3$ doped epitaxial polysilicon, with $E = 166000$ MPa and $v = 0,23$.

Width was imposed by the microfabrication process, being the thickness of the epitaxial polysilicon layer, as well as the distance between the microbeam and the silicon layer underneath located, being 4.1 μm, corresponds to the thickness of the sacrificial silicon oxide layer removed by etching. The massive structure of the electrode supporting the microbeam helped microfabrication process to obtain the longest beams by etching. In the electromechanical coupling it regularizes the electrostatic field across the gap. An optimised value of 2 μm of thickness was found as compromise between the need of a sufficient electrostatic actuation to bend the specimen and the electric breakdown. Length, width and gap values were selected to have a good variety of aspect ratios, as it is discussed in following paragraphs.

Connection pads, obtained by deposition of Aluminium alloy layer, offer a square contact area, whose side is 80 μm, to allow a stable contact to the probes used to apply the electrostatic actuation (Fig.2). A third pad, located on the edge of the die is connected to the silicon layer under the beam. Two connection pads have been placed on the counter-electrode, in order to keep the probes distant enough to avoid parasitic effects and mutual interference.

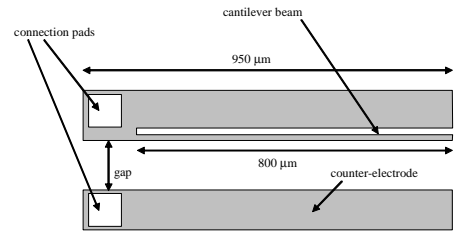

*Figure 2: Example of microcantilever.*

Geometrical dimensions of the microbeams were measured by means of Fogale Zoomsurf 3D. Some differences between the nominal and the actual values, due to the process tolerances, were detected, as shown in Table 1. Listed values include a range of variation of the parameters on the population of specimens, having equal geometry, up to seven microstructures. Experiments were performed by identifying the single specimen within the group having the same reference code, in Table 1. Code ST1 identifies this first set of specimens, while each layout is identified by the second number, like ST1-1. Measured length $l$, width $w$, thickness $t$, gap $g$ and aspect ratios from $R_1$ to $R_4$ are evidenced by symbol (*).

## 2.2. Remarks on the aspect ratios of microspecimens

Specimens were designed by taking care of four basic aspect ratios, which affect their mechanical behaviour:

$$R_1 = w/l \; ; R_2 = g/l \; ; R_3 = t/l \; ; R_4 = t/w \qquad (1)$$

In particular, $R_1$ may warn about the limit of application of beam model with respect to the plate's one [1]; $R_2$ foresees the possibility of large displacement [1], i.e. geometrical nonlinearity [14-18], while $R_3$ and $R_4$ are used to evaluate the beam stiffness, even to predict the anticlastic curvature [1]. Results show that specimens ST1-1, ST1-2, ST1-3 may need to resort to plate model. Width values may motivate a certain influence of the three dimensional nature of the electric field, affecting the actual value of the electromechanical force. The lateral curvature does not seem dominant, to require to include this deformation in the models. Specimens ST1-3, ST1-7, ST1-8 are prone to exhibit geometrical nonlinearity, caused by large tip displacement, if compared to the length of the beam. Fairly compliant are microbeams ST1-6;-7;-8.

## 2.3. Experimental set-up

Experimental validation was performed by the optical profiling system Fogale Zoomsurf 3D, based on non-contact optical interferometry [19]. The maximum lateral resolution is similar to that of the conventional optical microscopes (diffraction limited, 0.6μm with a 20X objective), while the vertical resolution may reach 0.1 nm. Optical magnification can reach up to 32X. The recorded light intensity is detected by a CCD pixel as function of the specimen height, thus defining either the profile of the monitored specimen or its position within the lighted area.

Microbeams were fabricated on square chips of 3 mm. To prevent any accidental motion of the chip, the latter was fixed on the motorized *XY* in-plane translation stage of the profiler, by a glass slide. The objective is equipped by a motorized *Z* translation stage, allowing the motion along the column, which is controlled.

| ID Ner | l | w | t | g | R1 | R2 | R3 | R4 | l* | | w* | | t* | | g* | | R1* | R2* | R3* | R4* |
|---|---|---|---|---|---|---|---|---|---|---|---|---|---|---|---|---|---|---|---|---|
| ST1-1 | 100 | 15 | 2 | 5 | 0,150 | 0,050 | 0,020 | 0,133 | 101,00 | ±0,1 | 15,00 | | 1,80 | ±0,02 | 5,00 | ±0,3 | 0,149 | 0,050 | 0,018 | 0,120 |
| ST1-2 | 100 | 15 | 2 | 10 | 0,150 | 0,100 | 0,020 | 0,133 | 101,00 | ±0,1 | 15,00 | | 1,80 | ±0,02 | 10,00 | ±0,3 | 0,149 | 0,099 | 0,018 | 0,120 |
| ST1-3 | 100 | 15 | 2 | 20 | 0,150 | 0,200 | 0,020 | 0,133 | 101,00 | ±0,1 | 15,00 | | 1,80 | ±0,02 | 20,10 | ±0,3 | 0,149 | 0,199 | 0,018 | 0,120 |
| ST1-4 | 200 | 15 | 2 | 10 | 0,075 | 0,050 | 0,010 | 0,133 | 205,00 | ±0,2 | 15,00 | | 1,90 | ±0,02 | 10,00 | ±0,3 | 0,073 | 0,049 | 0,009 | 0,127 |
| ST1-5 | 200 | 15 | 2 | 20 | 0,075 | 0,100 | 0,010 | 0,133 | 205,00 | ±0,2 | 15,00 | | 1,90 | ±0,02 | 20,00 | ±0,3 | 0,073 | 0,098 | 0,009 | 0,127 |
| ST1-6 | 800 | 15 | 2 | 40 | 0,019 | 0,050 | 0,003 | 0,133 | 805,00 | ±0,5 | 15,00 | | 2,70 | ±0,04 | 39,60 | ±0,3 | 0,019 | 0,049 | 0,003 | 0,180 |
| ST1-7 | 800 | 15 | 2 | 200 | 0,019 | 0,250 | 0,003 | 0,133 | 805,00 | ±0,5 | 15,00 | | 2,70 | ±0,04 | 200,00 | ±0,5 | 0,019 | 0,248 | 0,003 | 0,180 |
| ST1-8 | 800 | 15 | 2 | 400 | 0,019 | 0,500 | 0,003 | 0,133 | 805,00 | ±0,5 | 15,00 | | 2,70 | ±0,04 | 400,00 | ±0,5 | 0,019 | 0,497 | 0,003 | 0,180 |

*Table 1: Synoptic table of nominal and actual dimensions (\*) and related aspect ratios of set ST1 of in-plane bending microbeams (units [μm]).*

Microbeam is bended by the electromechanical action, induced by the electric field, when voltage is applied between the beam electrode and the counter-electrode, through the connecting pads. Power supplier is internal in Fogale Zoomsurf 3D and supplies only up to 200 Volt. Connection between power supplier and circuit was assured by adjustable needles, mounted on the ProbeHeads PH100 Suss. The latter have a mobile arm, with a pivot, which was magnetically fixed on the work plane of the instrument (Fig.3). The needle position was driven on the pad by means of three screws, controlling the motion along the three directions. Tests were performed by applying a positive voltage to the counter-electrode and connecting the beam, the electrode and the silicon wafer all together to the ground (null voltage). This configuration avoids unforeseen deflections of the microbeam under the bias voltage and minimizes the fringing field effect. Static deflection was detected by processing the high resolution images, obtained by white light measurement, through a scanning of the interferometric fringes on the focused area of the monitored specimen. In practice, scanning rectangle included the tip of the beam and part of the massive element of the electrode, as in Figure 4. Interferometric measurement provided a top view of the specimen, limited to the focused window, then a quoted profile of the transversal section of the microbeam (Fig.4).

### 3. THE EXPERIMENTAL VALIDATION

Models validation was based on the experimental reconstruction of the curve displacement vs. voltage, to verify, point by point, the correspondence of the actual tip displacement, measured by Zoomsurf 3D and the predicted numerical values. A geometrical linear solution, which assumes small displacement, is compared to the nonlinear approaches, implemented by means of the Finite Element Method (FEM).

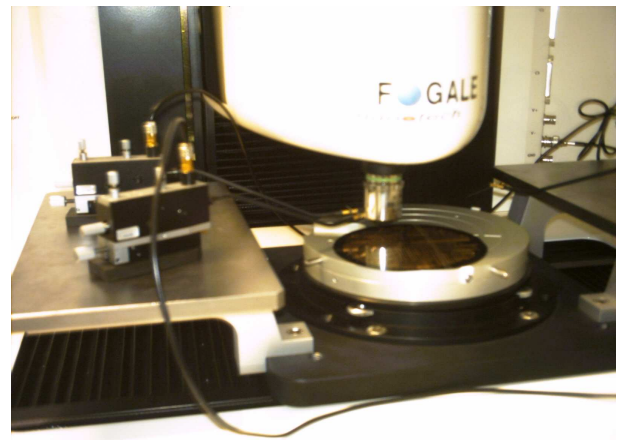

*Figure 3: Experimental set-up on Fogale Zoomsurf 3D.*

The electromechanical coupling between the electrical and mechanical degrees of freedom motivates to resort to a so-called coupled-field analysis, including both mechanical and electrical degrees of freedom. The original method proposed by the authors in [16-18], consists of a non incremental solution of the coupled problem, made possible by introducing a special finite beam element (so called SFET) suitable to operate even in presence of large displacement. This method is compared to the results of a coupled-field approach, based on a FEM iterative solution, which applies a morphing of the elements in the dielectric region, to avoid the effects of the element distortions. This approach is available in commercial code ANSYS, by meshing elements PLANE121 and PLANE183. All numerical outputs were compared to the experimental results in Figures 5, 6, 7. Plane models were initially implemented, to distinguish the effects of nonlinear electromechanical coupling, from those due to the three dimensional nature of the problem. A complete investigation about the differences between two and three dimensional models are currently carried out and validated. In practice, these investigations demonstrate the presence of local effects, in the electric field distribution, affecting the actual value of the electrostatic forces and somewhere the pull-in prediction.

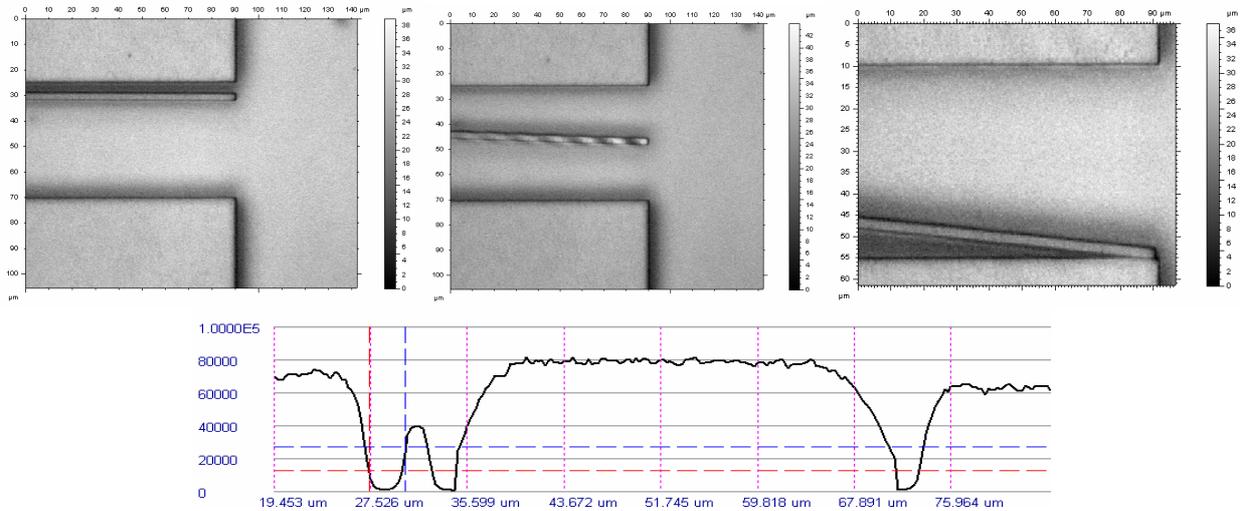

*Figure 4: Experimental images and profile provided by the Zoomsurf 3D Fogale.*

Results in terms of pull-in voltage were compared even to the analytical simplified solutions, computed by means of lumped parameters models proposed in [21]. For each set of specimens the validation has been completed, by experimentally measuring the tip displacement of the microcantilever, for a gradually increasing voltage, up to pull-in, every time it was allowed by the operating conditions. Moreover, the same measurement was repeated several times and averaged on the same sample, up to twelve times, depending on the occurrence of accidental failure or destructive pull-in.

### 3.1. Analytical approaches

A preliminary analysis was performed on pull-in voltage and related displacement to give a figure of the expected values on the experiments, by means of the well known formulas proposed in [21]. Results are immediately compared to the experimental evidences, where it was possible, in Table 2. Since several specimens exhibit pull-in voltage above the limit of 200 V of the Zoomsurf 3D power supplier, in absence of an external supplier, comparison was limited to the maximum value of voltage reached. Prediction of pull-in parameters looks quite good, although approximated, if performed according to Senturia-Osterberg formulation as [1, 21]:

$$V_{PI} = \sqrt{\frac{0.28 g_0^3 t^3 E}{\varepsilon l^4 \left(1 + 0.42 \frac{g_0}{w}\right)}}; \quad v_{PI} = \frac{3}{4}\frac{\varepsilon l^4}{E g_0^2 t^3} V_{PI}^2 \qquad (2)$$

where symbols mean: $g_0$ initial gap, $t$ thickness, $E$ Young's modulus, $\varepsilon$ dielectric constant, $l$ length, $V_{PI}$ pull-in voltage, $v_{PI}$ pull-in displacement.

### 3.2. FEM approaches

A complete experimental validation was performed on the set of specimens ST1, described in Table 1. The most relevant results are herewith summarized and compared in figures 5-7.

A first nonlinear and non incremental approach, suitable to predict even large displacement was implemented in MATLAB, according to [16-18]. FEM discretization was applied to consider only the most significant specimens ST1-1/4/6, mesh was generated as follows. The structure was described by 20 3-node SFET elements (Special beam element) [17, 18] for a total of 41 nodes, and for the dielectric 5633 nodes, with 2672 6-node isoparametric triangular finite elements for ST1-1, 5409 nodes and 2544 6-node isoparametric triangular finite elements for ST1-4 and 5301 nodes, 2472 6-node isoparametric triangular finite elements ST1-6. All the above models were implemented by the authors in MATLAB.

The latter method was compared to the iterative approach, including mesh morphing and geometrical nonlinear solution, implemented for instance in ANSYS, through PLANE183, 8-node isoparametric quadrilateral finite elements, solid (beam) and PLANE 121 8-node isoparametric quadrilateral finite elements (electrostatic).

In this case a suitable mesh consisted of 80 beam elements and started with 9500 nodes and 3000 elements, but the number of PLANE121 elements was updated during the computation by the code, through a re-meshing operation, or "morphing", applied to the geometrical nonlinear solution.

|  | Analytical | | Experimental | |
|---|---|---|---|---|
| N. | Voltage [V] | Displ. [µm] | Voltage [V] | Displ. [µm] |
| ST1-1 | 180 | 0,92 | 184 | 1,6 |
| ST1-2 | 480 | 1,64 | N.A. | N.A. |
| ST1-3 | 1253 | 2,69 | N.A. | N.A. |
| ST1-4 | 126 | 1,64 | 136 | 3 |
| ST1-5 | 323 | 2,70 | N.A. | N.A. |
| ST1-6 | 86 | 3,92 | crashed | crashed |
| ST1-7 | 546 | 6,36 | N.A. | N.A, |
| ST1-8 | 1137 | 6,88 | N.A. | N.A, |

*Table 2: Validation of the analytical model on pull-in.*

Since Young modulus of the material was known only approximately, some measures were performed through a dynamic response of the microcantilevers [1, 12, 21]. Results showed a certain variability of the values, therefore a minimum and a maximum value of 150 GPa and 166 GPa respectively were inputted into the simulation to investigate the model sensitivity on this parameter, and results were drawn in Figures 5-7.

## 4. DISCUSSION

The influence of the geometrical nonlinearity due to the large displacement of the tip of the tested microcantilevers is sufficiently high to motivate the implementation of a nonlinear structural and coupled analysis. As figures 5 and 6 show, the behavior close to pull-in condition becomes nonlinear and differences with linear solution are remarkable. Specimen ST1-6 exhibits the same behavior, but the accidental failure of the specimens did not allow to reach the pull-in. Value of Young modulus affects the computation of pull-in, but not significantly like the thickness. Experiments show that nominal values of *E* never fitted the actual response of the structure, but all results are enclosed in the area delimited by the curves computed with the two selected values. Results of the nonlinear model based on SFET element are consistent with the actual behavior of the specimens, although the value of pull-in voltage is always predicted with a certain approximation. The coupled field approach, with morphing, based on elements PLANE121/183, overestimated a little bit the actual behavior in above tests.

| Legend | | | |
|---|---|---|---|
| —— Linear (166 GPa) | | O | Experiments |
| – – Non incremental (150 GPa) | | —●— | (166 GPa) |
| • PLANE121 / 183 (150 GPa) | | + | (166 GPa) |

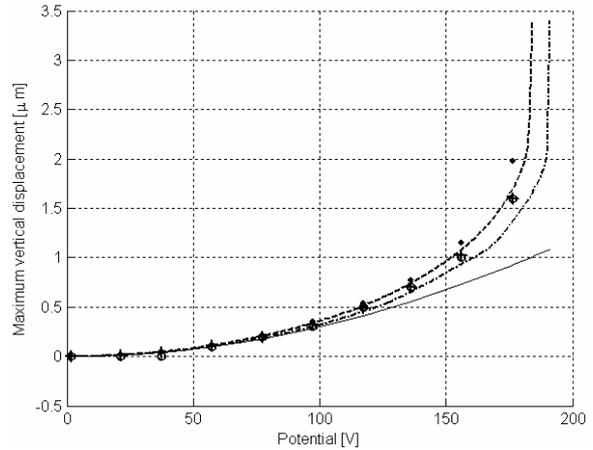

*Figure 5: Comparison for specimen ST1-1.*

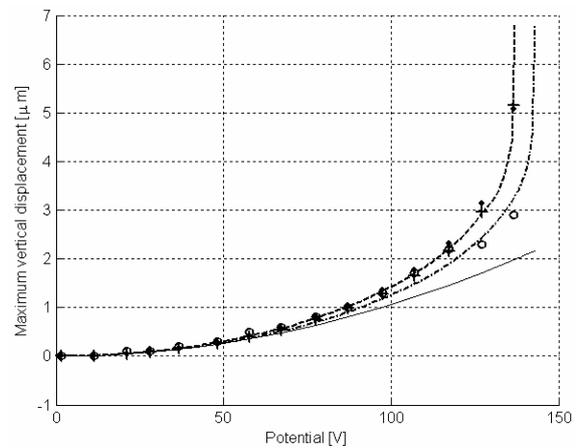

*Figure 6: Comparison for specimen ST1-4.*

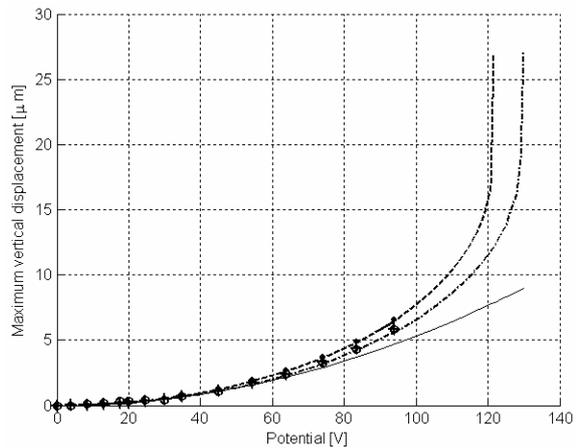

*Figure 7: Comparison for specimen ST1-6.*

Nevertheless it validated and confirmed the prediction of the proposed FEM approach [16-18]. Computational times are comparable for these two FEM solutions. Authors are currently investigating the possibility to enhance the performances of the solution algorithms by means of a discretization based on mixed methods not only FEM-BEM, but also on the Cell Method.

## 5. CONCLUSIONS

This study was aimed to validate the numerical approaches proposed in [14-18] to predict the static deflection of microbeams electrostatically actuated. Experiments demonstrated that, even for small values of aspect ratios described in Table 1, the geometrical nonlinearity, mainly due to large displacements in microcantilevers is influent and should be implemented into the numerical design tools. The non incremental approach [17, 18] gives results consistent with the experiments, and compatible with the coupled-field, nonlinear and iterative solution available in commercial codes like ANSYS. Computational time is comparable for the two FEM approaches. Current activity carried by the authors concerns some additional effects, due to the three dimensional nature of the electric field, a double clamped layout and residual stresses, and the discretization of the dielectric field by Cell Method to enhance the computational performance.

## 7. AKNOWLEDGEMENTS

This work was partially funded by the Italian Ministry of University, under grant PRIN-2005/2005091729, while microspecimens were built by STMicroelectronics (Cornaredo, Italy).